\newcommand{\spitzer}{\textit{Spitzer}}

\newcommand{\lsim}{\lesssim}

\newcommand{\gsim}{\gtrsim}
\newcommand{\lstar}{\hbox{$L^\ast$}}
\newcommand{\mcal}{\hbox{$\mathcal{M}$}}

\newcommand{\msol}{\hbox{$\mathcal{M}_\odot$}}

\newcommand{\lsol}{\hbox{$L_\odot$}}

\newcommand{\lir}{\hbox{$L_{\mathrm{IR}}$}}

\newcommand{\micron}{\hbox{$\mu$m}}
\newcommand{\mnras}{\textit{Mon.\ Not.\ Royal.\ Astro.\ Soc.}}
\newcommand{\aap}{\textit{Astron.\ \& Astrophys.}}
\newcommand{\aj}{\textit{Astronom.\ J.}}
\newcommand{\apj}{\textit{Astrophys.\ J.}}
\newcommand{\apjl}{\textit{Astrophys.\ J.\ Lett.}}
\newcommand{\apjs}{\textit{Astrophys.\ J.\ Sug.}}
\newcommand{\araa}{\textit{Ann.\ Rev.\ Astron.\ \& Astrophys.}}

\newcommand{\mysection}[1]{\vspace{10pt} \noindent\textbf{#1}}

\documentclass{iaus}
\usepackage{graphicx}

\title[The Assembly of High Redshift Galaxies] 
{The Star Formation History and Stellar Assembly of High Redshift Galaxies}

\author[C.~Papovich]   
{Casey Papovich
}
\affiliation{Spitzer Fellow, Steward Observatory, 933 North Cherry
Avenue, Tucson, Arizona, 85721, USA \break email: papovich@as.arizona.edu}

\pubyear{2006}
\volume{235}  
\pagerange{1-4}
\date{}
\jname{Galaxies Across the Hubble Time}
\editors{J.~Palous \& F.~Combes, eds.}
\begin{document}

\maketitle

\begin{abstract}
I discuss current observational constraints on the star-formation and
stellar-assembly histories of galaxies at high redshifts.
The data on massive galaxies at $z<1$ implies that their stellar
populations formed at $z>2$, and that their morphological
configuration was in place soon thereafter.   \spitzer\ Space Telescope
24~\micron\ observations indicate
that a substantial fraction of massive galaxies at $z\sim 1.5-3$ have
high IR luminosities, suggesting they are rapidly forming stars,
accreting material onto supermassive black holes, or both.   I compare
how observations of these IR--active phases in the histories of
massive galaxies constrain current galaxy--formation models.
\end{abstract}


\noindent Most ($\sim$50\%) of the stellar mass in galaxies today
formed during the short time between $z$$\sim$3 and 1 (e.g., Dickinson
et al.\ 2003, Rudnick et al.\ 2006).   Much of this stellar mass
density resides in massive galaxies, which appear at epochs prior to
$z$$\sim$1--2 (see, e.g., McCarthy 2004, Renzini 2006).   The
fashionable scenario is that galaxies ``downsize'', with massive
galaxies forming most of their stars at early cosmological times, with
less--massive galaxies continuing to form stars to the present (e.g.,
Juneau et al.\ 2005).   However, it is still unclear when and where
the stars in these galaxies formed.  For example, it may be that stars
form predominantly in low--mass galaxies at high redshifts, which then
assemble over time to form large, present--day massive galaxies (e.g.,
Kauffmann \& Charlot 1998).

At $z$$\lsim$1 massive galaxies exist on a fairly prominent red
sequence (e.g., Blanton et al.\ 2003, Bell et al.\ 2004, Willmer et
al.\ 2006), are largely devoid of star formation and evolve passively.
Forming such a red sequence is a challenge for contemporary
hierarchical galaxy formation models (e.g., Dav\'e et al.\ 2005)
without including some agent to suppress star formation (the favorable
mechanism is feedback from AGN; e.g., Croton et al.\ 2006, Hopkins et
al.\ 2006).   In hierachical models, the most massive galaxies
continue to grow via satellite accretion.  To maintain the red
sequence this growth must occur without continued star--formation
(so--called ``dry'' merging), but so far observational evidence is
inconclusive and conflicting  (e.g., van Dokkum
2005; Bell et al.\ 2006;  Faber et al.\ 2005).  

Understanding galaxy formation boils down to two questions:  When did
galaxies form their stars?   And, when did they assemble into their
present--day configurations?   In these proceedings, I discuss
constraints on star formation in massive galaxies at
$z$$\sim$1.5--3. In particular I focus on observations of the IR
activity in high redshift galaxies using \spitzer\ 24~\micron\
observations, and what this means for the galaxies' assembly and
evolution.

\mysection{Star Formation in High--$z$ Massive Galaxies}

\noindent  Deep \spitzer\ surveys at 24~\micron\ show that
IR luminous galaxies evolved very rapidly (Papovich et al.\ 2004),
dominating the SFR density at $z\sim 1$ (e.g., Le Floc'h et al.\
2005).    Several studies of the 24~\micron\ emission of higher
redshift galaxies ($1.5 < z < 3$) show very high detection rates
($\gsim 50$\%; Daddi et al.\ 2005; Papovich et al.\ 2006; Reddy et
al.\ 2006; Webb et al.\ 2006), suggesting that the majority of
high--redshift galaxies emit in the thermal IR --- \textit{they are
either actively forming stars, supermassive blackholes, or both at
this epoch}.   
%
\begin{figure}[t]
\centering\includegraphics[width=2.5in, height=2.5in]{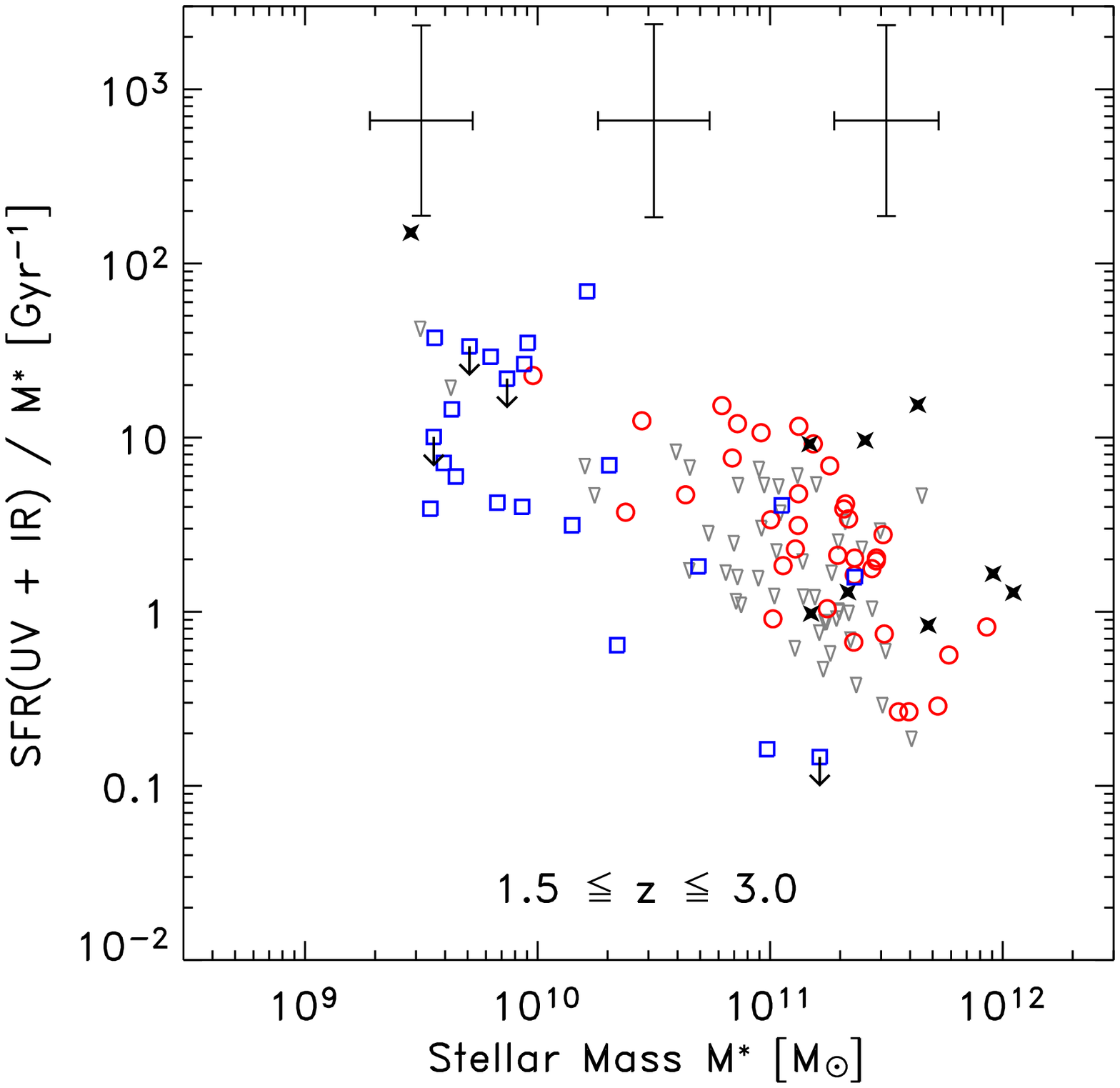}
\includegraphics[width=2.5in, height=2.5in]{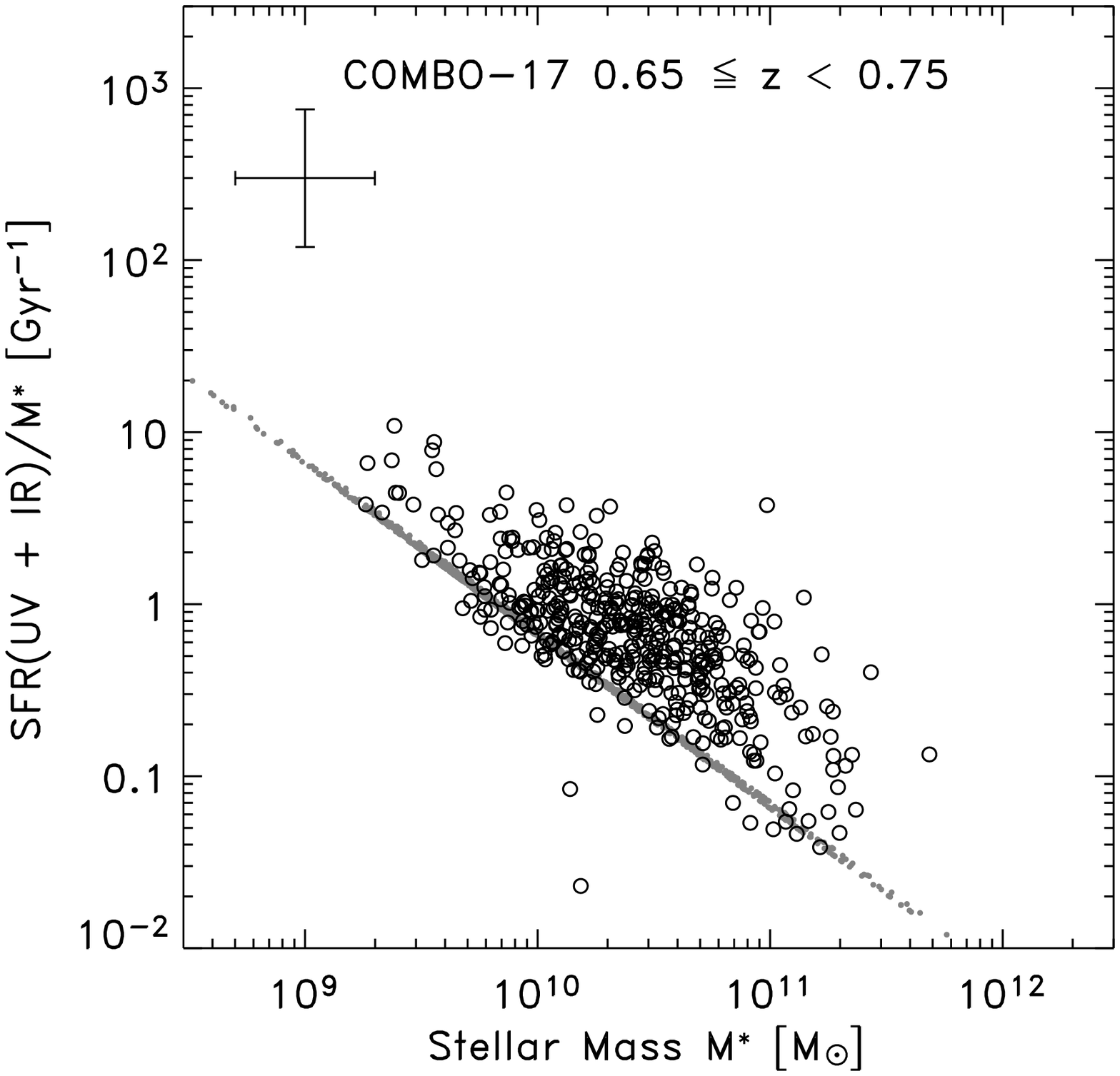}
\vspace{-12pt}
\caption{Specific SFR as a function of stellar mass for high redshift
galaxies (from Papovich et al.\ 2006).   The left panel shows galaxies
at $1.5 < z < 3.0$.  Red circles correspond to DRGs; gray triangles
denote SFR upper limits.  Blue squares show galaxies from the HDF--N.
Stars show X--ray sources.  The right panel shows galaxies from
COMBO--17 at $0.65 < z < 0.75$.  Dots denote SFR upper limits.}
\end{figure}

For example, figure~1 shows the specific SFRs ($\Psi/\mcal$) derived
from the masses and SFRs for the galaxies in the GOODS fields at $1.5
< z < 3.0$ (from Papovich et al.\ 2006), where the SFRs are derived
from the summed UV and IR emission based on the \spitzer\ 24~\micron\
data.   The figure also shows the specific SFRs for $0.65 < z
< 0.75$ galaxies from COMBO--17 (Wolf et al.\ 2003), derived also
using \spitzer\ 24~\micron\ data.  Interestingly there is strong
evolution in the specific SFRs, especially for massive galaxies.
Galaxies with masses $\geq$$10^{11}$~\msol\ at 1.5$\leq$$z$$\leq$3
have high specific SFRs,  $\Psi/\mcal$$\sim$0.2--10 Gyr$^{-1}$,
excluding X--ray sources.   In contrast, at $z$$\lsim$ 0.75 galaxies
with $\mcal$$\geq$$10^{11}$~\msol\ have much lower specific SFRs,
$\Psi/\mcal$$\sim$0.1--1 Gyr$^{-1}$.

\begin{figure}
\centering\includegraphics[width=3.5in,height=3.0in]{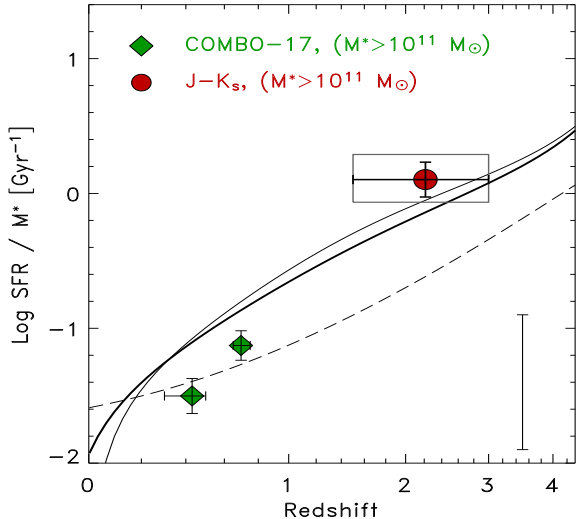}
\vspace{-12pt}
\caption{The integrated specific SFR as a function of redshift (from
Papovich et al.\ 2006).  The integrated specific SFR is the ratio of
the sum of galaxy SFRs to the sum of galaxy stellar masses.   The
curves show the expected evolution from the global SFR density.  Data
points show results for galaxies with $\geq$$10^{11}$ \msol.   The
inset error bar shows an estimate on the systematics.}
\end{figure}

Papovich et al.\ (2006) defined the integrated specific SFR as the
ratio of the sum of all galaxy SFRs, $\Psi$, to the sum of their
stellar masses, $\mcal$.   Figure~2 shows the integrated specific SFRs
for galaxies selected from GOODS  at $z$$\sim$1.5--3.0 and COMBO--17
at $z\sim 0.4$ and 0.7, all with $\mcal \geq 10^{11}$~\msol.   The
error box indicates the affect of assumptions in the SFRs and AGN
activity  (see further discussion below, and in Papovich et al.\
2006).   The integrated specific SFR in galaxies with
\mcal$>$$10^{11}$~\msol\ declines by more than an order of magnitude
from $z$$\sim$1.5--3 to $z$$\lsim$0.7.  The curves in figure~2 show
the specific SFR integrated over all galaxies, not just the most
massive;  this is the ratio of the cosmic  SFR density to its
integral, $\dot{\rho}_\ast / \int \dot{\rho}_\ast\, dt$.   Although
there is a decrease in the global specific SFR with decreasing
redshift, the  evolution in the integrated specific SFR in massive
galaxies is accelerated.   The implication is that at $z$$\gsim$1.5,
massive galaxies are rapidly forming their stars, whereas by
$z$$\lsim$1.5 the specific SFRs of massive galaxies drops rapidly, and
lower--mass galaxies dominate the cosmic SFR density.

If AGN contribute to the observed 24~\micron\ emission in galaxies at
$z$$\sim$1.5--3, then they can affect the inferred IR luminosities.
For example, using an IR template for Mrk~231 instead of a
star--forming galaxy with $\lir$$\gsim$$10^{13}$~\lsol\ would reduce
the IR luminosity for $z$$\sim$1.5--3 galaxies by a factor of
$\sim$2--5.   Many ($\sim$15\%) of the massive galaxies at
$z$$\sim$1.5--3 are detected in the deep X--ray data (see Papovich et
al.\ 2006), and these objects tend to have high inferred specific SFRs
(see figure~1).   This obviously has an effect on the evolution of the
integrated specific SFRs:  the error box on the data point at $1.5 < z
< 3$ in figure~2 shows how the result changes if the SFR for galaxies
with putative AGN (based on X--ray detections, or rest--frame near--IR
colors, see Stern et al.\ 2005; Alonso--Herrero et al.\ 2006) is set
to zero.  Interestingly, the high AGN occurrence in massive galaxies
suggests that at $z$$\sim$1.5--3 these objects are forming simultaneously
their stars and supermassive black holes.  This may provide the
impetus for the present--day black-hole--bulge-mass relation and/or
provide the feedback necessary to squelch star--formation in such
galaxies, moving them onto the red sequence.

\mysection{Confrontation with Models}

\noindent Recent hierarchical galaxy--formation models predict a
``downsizing'' trend in the star formation rates of massive galaxies.
De Lucia et al.\ (2006) show that within the semi-analytical
galaxy--formation prescription coupled with feedback from AGN (Croton
et al.\ 2006) that galaxies in the most massive dark matter haloes
formed their stars at the earliest epochs.   These
models seem broadly consistent with observations. The ``downsizing'' jargon
used by astronomers  merely signifies that star--formation is accelerated in the most
massive overdensities in current $\Lambda$CDM
galaxy--formation models.

While encouraging, the details of star--formation in massive galaxies
are not yet fully consistent with the \spitzer\ 24~\micron\
observations.   Assuming the galaxies at $1.5 < z < 3$
with observed masses $>10^{11}$~\msol\ evolve to present--day galaxies
with masses of \textit{at least} this much, then the De Lucia et al.\
model predicts they should have observed specific SFRs
$\sim$0.3~Gyr$^{-1}$ on average.  The observations instead suggest a
specfic SFR value closer to 1~Gyr$^{-1}$ (see figure~1).  There may be
a discrepancy at the factor $\sim$3 level, and this is likely
discrepancy is likely greater because the observed galaxies presumably
will continue to increase their stellar mass to $z$$\sim$0.
Admittedly the uncertainties of the IR--infered SFRs are at the factor
$\sim$3 level, and to provide constraints on models of the star formation
rates of massive galaxies will require lowering these observational
uncertainties.

A more serious challenge to hierarchical models is the existence of
a substantial population of apparently  passive galaxies on the red
sequence by $z\sim 1$.   The stellar mass density of galaxies on
the red sequence is near its present--day value by $z\sim 0.7$ (e.g.,
Brown et al.\ 2006).  Furthermore, Cimatti et al.\ (2006; see also Cimatti et al.,
\textit{these proceedings}) find that by using empirical
color--evolution models the number of the most luminous galaxies
($>$4~\lstar) on the red sequence is nearly unchanged to $z\sim 1$.
This is difficult for hierarchical models that predict the most
massive present--day galaxies assembled into their current
configurations at the lowest redshifts (e.g., Neistien et al.\ 2006;
De Lucia et al.\ 2006).   For example, both the Neistein et al.\ and
De Lucia et al.\ show that the main progenitor (having $\gsim 50$\% of
the final mass) of massive galaxies with present--day $M >
10^{12}$~\msol\ formed at $z<<1$.   This is difficult to reconcile
even with frequent ``dry merging'' on the red sequence (see also Faber
et al.\ 2005).

We are witnessing a growing understanding (and even possible
convergence)  between observational and theoretical constraints on
galaxy evolution, even though significant hurdles remain.  It will be
exciting to see the summary of our knowledge in this field at the next
IAU General Assembly.

\begin{acknowledgments}
\noindent I wish to thank the IAU symposium organizers for the invitation to
present this material, and for planning a successful meeting in
such a historic locale.   I am grateful for my colleagues on the MIPS
GTO and GOODS teams for their continued collaboration.   Support for
this work was provided by NASA through the Spitzer Space Telescope
Fellowship Program, through a contract issued by JPL/Caltech under a
contract with NASA.
\end{acknowledgments}


%

\end{document}